# Interference control of perfect photon absorption in cavity quantum electrodynamics

Liyong Wang[1-4], Ke Di[1], Yifu Zhu[1], and G. S. Agarwal[5-6]

[1]Department of Physics, Florida International University, Miami, Florida 33199, USA
[2]Center for cold atom physics, Chinese Academy of Sciences, Wuhan, 430071, China
[3]Wuhan Institute of Physics and Mathematics, Chinese Academy of Sciences, Wuhan, 430071, China
[4]University of Chinese Academy of Sciences, Beijing 100049, China 230026, China
[5]Department of Physics, Oklahoma State University, Stillwater, Oklahoma 74078, USA
[6]Institute for Quantum Science and Engineering, and Department of Biological and Agricultural Engineering, Texas A&M University, TAMU4242, College Station, TX 77843

We propose and analyze a scheme for controlling coherent photon transmission and reflection in a cavity-quantum-electrodynamics (CQED) system consisting of an optical resonator coupled with three-level atoms coherently prepared by a control laser from free space. When the control laser is off and the cavity is excited by two identical light fields from two ends of the cavity, the two input light fields can be completely absorbed by the CQED system and the light energy is converted into the excitation of the polariton states, but no light can escape from the cavity. Two distinct cases of controlling the perfect photon absorption are analyzed: (a) when the control laser is tuned to the atomic resonance and creates electromagnetically induced transparency, the prefect photon absorption is suppressed and the input light fields are nearly completely transmitted through the cavity; (b) when the control laser is tuned to the polariton state resonance and inhibits the polariton state excitation, the perfect photon absorption is again suppressed and the input light fields are nearly completely reflected from the cavity. Thus, the CQED system can act as a perfect absorber or near perfect transmitter/reflector by simply turning off or on of the control laser. Such interference control of the coherent photon-atom interaction in the CQED system should be useful for a variety of applications in optical logical devices.

I.   INTRODUCTION

Studies of radiation and matter interactions continue to play an essential role in the advancement of fundamental physics and development of practical applications. It is desirable to be able to manipulate and control the light absorption, emission, and scattering in an optical medium. One widely used technique for such studies in recent years is electromagnetically induced transparency (EIT), with which the light transmission through an absorbing medium can be enhanced [1-2]. Recently it has been shown that the light fields coupled to an optical cavity with an intra-cavity absorber can be completely absorbed and no light can escape through the cavity [3-8].  Such a cavity-absorber system acts as a coherent perfect absorber and may be useful for a variety of fundamental studies and practical applications [9-19].

Here we propose a scheme for controlling the coherent photon-atom interactions by combining the atomic interference technique with the perfect photon absorption in a cavity quantum electrodynamics (CQED) system. The scheme enables the CQED system to act as a perfect coherent absorber [20] or a near perfect transmitter/reflector manifested by the quantum interference induced by a control laser coupled to the intra-cavity atoms from free space.

Specifically, the proposed CQED system consists of a cavity containing three-level atoms and is excited by two coherent light fields from two output mirrors. A control laser is coupled to the atoms from the open free-space side of the cavity. In section II, we present the theoretical model and derive analytical results for the CQED system. We specify the conditions for the prefect photon absorption, under which when the control laser is off, the photons from the two input light fields coupled into the cavity are completely absorbed through the excitation of the CQED polariton state and there is no output light from the cavity. We show that the perfect photon absorption in the CQED system can be understood in terms of a simple one-sided cavity model with matched photon loss rate from the mirror and other intra-cavity photon loss rate. This is analogous to an impedance-matched AC electric circuit in which the load power is maximized and there is no power reflection from the circuit. In section III, we show that the CQED system can be made to act as a perfect photon absorber (without the control laser) or as a near perfect photon transmitter (when the control laser is on and creates the cavity EIT [21-25]). In section IV, we show that when the control laser is tuned to the polariton resonance and suppresses the polariton excitation, the input light fields cannot be coupled into the cavity and are nearly completely reflected from the cavity. Thus the control laser can switch the CQED system from a perfect photon absorber to a near perfect photon reflector. In section V, we present more detailed numerical calculations and characterize the switching of the CQED system from a perfect absorber to a near perfect reflector. Finally the summary and conclusion is presented in section VI.

## II. COHERENTLY PREPARED CQED SYSTEM WITH PERFECT PHOTON ABSORPTION

Fig. 1 shows the schematic diagram for the coupled CQED system that consists of N three-level atoms confined in a single mode cavity and is excited by two input light fields $a_{in}^r$ and $a_{in}^l$ from two ends, and a control field from free space. The cavity mode couples the atomic transition |1>-|3> with frequency detuning $\Delta_c = \omega_c - \omega_{31}$. The two input fields have the same frequency $\omega_p$ and are tuned from the atomic transition by $\Delta_p = \omega_p - \omega_{31}$. The control laser drives the atomic transition |2>-|3> with Rabi frequency $2\Omega$ ($\Omega = \frac{E\mu_{23}}{\hbar}$, E is the control field amplitude and $\mu_{23}$ is the transition dipole moment between states |2> and |3>). $\Delta = \omega - \omega_{23}$ is the control frequency detuning. It will be shown that under appropriate conditions, when the control laser is off, the CQED system behaves like a simple cavity QED system with two-level atoms and the perfect photon absorption can be observed in the strong collective-coupling regime, in which the CQED system acts as a perfect absorber and there is no output light from the cavity [20]; when the control light is on and induces either the cavity EIT [21-25] or the polariton-suppression interference [26], the perfect photon absorption is suppressed and the CQED system acts as a near perfect light transmitter or light reflector to the two input fields.

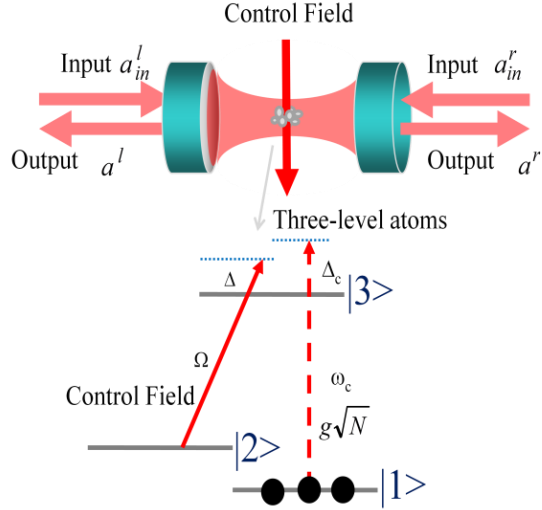

Fig. 1 The CQED system consisting of N three-level atoms confined in the cavity mode. Two input light fields are coupled into the cavity and a free-space control laser interacts with the atoms from the open side of the cavity.

In the rotating frames, the interaction Hamiltonian for the coupled CQED system under the rotating-wave approximation is

$$H_I = -\hbar(\sum_{i=1}^{N}\Omega\hat{\sigma}_{32}^{(i)} + \sum_{i=1}^{N}g\hat{a}\hat{\sigma}_{31}^{(i)}) + ia^+(\sqrt{2\kappa_r}a_{in}^r + \sqrt{2\kappa_l}a_{in}^l) + H.C. ,\qquad(1\text{-}1)$$

and the Hamiltonian for the free atoms and the cavity field is

$$H_{a+c} = -\hbar\sum_{i=1}^{N}\Delta_p\hat{\sigma}_{33}^{(i)} - \hbar\sum_{i=1}^{N}(\Delta_p - \Delta)\hat{\sigma}_{22}^{(i)} - \hbar(\Delta_p - \Delta_c)a^+a .\qquad(1\text{-}2)$$

Here $\hat{a}$ ($\hat{a}^+$) is the annihilation (creation) operator of the cavity photons, $a_{in}^l$ and $a_{in}^r$ are two input fields to the cavity (see Fig. 1), $\kappa_i = \dfrac{T_i}{\tau}$ (i=r or l) is the loss rate of the cavity field on the mirror i ($T_i$ is the mirror transmission and $\tau$ is the photon round trip time inside the cavity) and $\hat{\sigma}_{lm}^{(i)}$ (l, m=1-3) is the atomic operator for the ith atom. $g = \mu_{13}\sqrt{\omega_c/2\hbar\varepsilon_0 V}$ is the CQED coupling coefficients and is assumed to be uniform for the N identical atoms inside the cavity(thus we take $\hat{\sigma}_{lm}^{(i)} = \hat{\sigma}_{lm}$). We drop the quantum fluctuation terms and treat $\sigma_{lm}$ and $a$ as c numbers, then the equations of motion are

$$\dot{\sigma}_{11} = \gamma_{31}\sigma_{33} - iga\sigma_{31} + iga^*\sigma_{13} \qquad(2\text{-}1)$$

$$\dot{\sigma}_{22} = \gamma_{32}\sigma_{33} - i\Omega\sigma_{32} + i\Omega^*\sigma_{23} \qquad(2\text{-}2)$$

$$\dot{\sigma}_{33} = -(\gamma_{31}+\gamma_{32})\sigma_{33} + iga\sigma_{31} - iga^*\sigma_{13} + i\Omega\sigma_{32} - i\Omega^*\sigma_{23} \qquad(2\text{-}3)$$

$$\dot{\sigma}_{12} = -(\gamma_{21} - i(\Delta_p - \Delta))\sigma_{12} + i\Omega^*\sigma_{13} - iga\sigma_{32} \qquad(2\text{-}4)$$

$$\dot{\sigma}_{13} = -(\frac{\gamma_{31}+\gamma_{32}}{2} - i\Delta_p)\sigma_{13} - iga(\sigma_{33}-\sigma_{11}) + i\Omega\sigma_{12} \qquad (2\text{-}5)$$

$$\dot{\sigma}_{23} = -(\frac{\gamma_{31}+\gamma_{32}}{2} - i\Delta)\sigma_{23} - i\Omega(\sigma_{33}-\sigma_{22}) + iga\sigma_{21} \qquad (2\text{-}6)$$

$$\dot{a} = -((\kappa_1+\kappa_2)/2 - i(\Delta_p-\Delta_c))a + igN\sigma_{13} + \sqrt{\kappa_1/\tau}\,a_{in}^r + \sqrt{\kappa_2/\tau}\,a_{in}^l \qquad (2\text{-}7)$$

We assume $\gamma_{31}+\gamma_{32}=2\Gamma_3$ ($\Gamma_3$ is the decay rate of the excited state $|3\rangle$) and a symmetric cavity such that $\kappa_1=\kappa_2=\kappa$. $\gamma_{21}$ is the decoherence rate between the ground states $|1\rangle$ and $|2\rangle$ and typically much smaller than other decay rates. Then, with weak input light fields and under the condition g<<Ω, the atomic population is concentrated in $|1\rangle$ ($\sigma_{11}\approx 1$) and the steady-state solutions of Eq. (2) can be readily obtained analytically. The steady-state solution of the intra-cavity light field $a$ is then given by

$$a = \frac{\sqrt{\kappa/\tau}(a_{in}^r + a_{in}^l)}{\kappa - i(\Delta_p-\Delta_c) + \dfrac{g^2 N}{\Gamma_3 - i\Delta_p + \dfrac{\Omega^2}{\gamma_{12}-i(\Delta_p-\Delta)}}}, \qquad (3)$$

The steady-state solutions of the output light field from the right mirror and the left mirror are

$$a^r = \frac{\kappa(a_{in}^r + a_{in}^l)}{\kappa - i(\Delta_p-\Delta_c) + \dfrac{g^2 N}{\Gamma_3 - i\Delta_p + \dfrac{\Omega^2}{\gamma_{12}-i(\Delta_p-\Delta)}}} - a_{in}^r, \quad \text{and} \qquad (4\text{-}1)$$

$$a^l = \frac{\kappa(a_{in}^r + a_{in}^l)}{\kappa - i(\Delta_p-\Delta_c) + \dfrac{g^2 N}{\Gamma_3 - i\Delta_p + \dfrac{\Omega^2}{\gamma_{12}-i(\Delta_p-\Delta)}}} - a_{in}^l, \qquad (4\text{-}2)$$

respectively. If the two input fields are identical, $a_{in}^r = a_{in}^l$, the two output fields are equal, $a^r = a^l = a_{out}$. First the perfect photon absorption can occur in the CQED system when the control laser is off (Ω=0), in which the output light fields are zero ($a^r = a^l = 0$), but the intra-cavity light field $a \neq 0$. The required conditions for the perfect photon absorption are [20]

$$\frac{\Delta_p - \Delta_c}{\Delta_p} = \frac{\kappa}{\Gamma_3}, \qquad (7)$$

and

$$(\Delta_p - \Delta_c)\Delta_p = g^2 N - \kappa\Gamma_3. \qquad (8)$$

It has been shown that the prefect photon absorption in the CQED system can be readily achieved when the CQED system is in the strong collective coupling regime, $g^2 N \geq \kappa\Gamma_3$ [20]. Here we show that a CQED system with the prefect photon absorption is equivalent to a one-side cavity with matched photon loss rate κ/2 from the lossy mirror and the other intra-cavity photon losses $\Gamma$ (such as the photon scattering loss). As shown in Fig. 2 for a one-side cavity, the cavity light field $a$ in the steady state is given by

$$a = \frac{\sqrt{\kappa/\tau}\, a_{in}^l}{\kappa/2 + \Gamma + i\Delta_c} \qquad (9)$$

The only output field is from the left mirror and is given by

$$a_{out} = \frac{\kappa a_{in}^l}{\kappa/2 + \Gamma + i\Delta_c} - a_{in}^l \qquad (10)$$

When the input light field is resonant with the cavity, $\Delta_c=0$, the cavity output field $a_{out}=0$ if $\Gamma=\kappa/2$. That is, if the photon decay rate from the lossy mirror matches all other photon losses in the cavity, the input light field is completely coupled into the cavity, but there is no output light from the cavity mirror. Therefore, a CQED system with the perfect photon absorption is equivalent to a one-sided cavity with matched photon loss rates from the cavity mirror and other intra-cavity losses. This is analogous to the impedance match in an AC electric circuit, under which the load power is maximized and there is no reflection from the circuit back to the source.

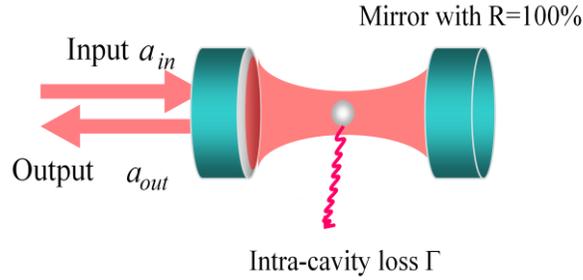

Fig. 2 One-sided cavity with perfect reflectivity of the right mirror (R=100%). An input field is coupled into the cavity from the left mirror. When the cavity photon loss rate κ (from the left mirror) is equal to the combined rate $\Gamma$ of all other intra-cavity photon losses, the input light field is completely coupled into the cavity and the output light field $a_{out}=0$.

Next we show that with a control laser, one can manipulate and change the characteristics of the perfect photon absorption and render the system useful for high-contrast all-optical switching applications. With $a_{in}^r = a_{in}^l = a_{in}$ (the two input fields are identical), and assume the cavity is tuned to the atomic resonance $\Delta_c=0$ and the system satisfies the perfect photon absorption condition ($\kappa=\Gamma_3$), we analyze two specific operations of the CQED system, one is controlled by the cavity EIT [21-25] and the other is controlled by the polariton-suppression interference [26].

## III. CAVITY EIT INDUCED SWITCHING FROM PERFECT PHOTON ABSORPTION TO NEAR PERFECT PHOTON TRANSMISSION

Under the perfect photon absorption condition, $\kappa=\Gamma_3$, we set $g^2N = \kappa\Gamma_3$ (the CQED system is at the threshold of the strong collective coupling regime). With $\Delta p=0$ and $\Delta=0$, the intra-cavity light field from Eq. (3) becomes

$$a = \frac{2\sqrt{\kappa/\tau}\, a_{in}}{\kappa[1+\kappa/(\kappa+\Omega^2/\gamma_{12})]}, \tag{11}$$

and the output light field of the cavity, $a^l = a^r = a_{out}$ and is

$$a_{out} = \frac{\Omega^2 a_{in}}{2\kappa\gamma_{12}+\Omega^2}. \tag{12}$$

When the control laser is off ($\Omega=0$), the intra-cavity light field is $a = \frac{a_{in}}{\sqrt{\kappa\tau}}$, but the output light field of the cavity is $a_{out}=0$. That is, the input light fields are completely absorbed and there is no output light from the cavity: the CQED system acts as a perfect photon absorber. When the control laser is turned on and its Rabi frequency $\Omega$ satisfies the condition, $\Omega^2 \gg 2\gamma_{12}\kappa$, the intra-cavity light field becomes $a = \frac{2a_{in}}{\sqrt{\kappa\tau}}$ and the output cavity field becomes $a_{out} = \frac{a_{in}}{1+\frac{2\kappa\gamma_{12}}{\Omega^2}} \approx a_{in}$. That is, the input light field with $\Delta p=0$ is nearly completely transmitted through the cavity. If one defines the on/off contrast of the cavity output intensity manifested by the control laser as $C = \frac{(|a_{out}|^2)_{control\text{-}on}-(|a_{out}|^2)_{control\text{-}off}}{|a_{in}|^2}$, then, $C = |\frac{1}{1+\frac{2\kappa\gamma_{12}}{\Omega^2}}|^2 \approx 1$, which is nearly perfect.

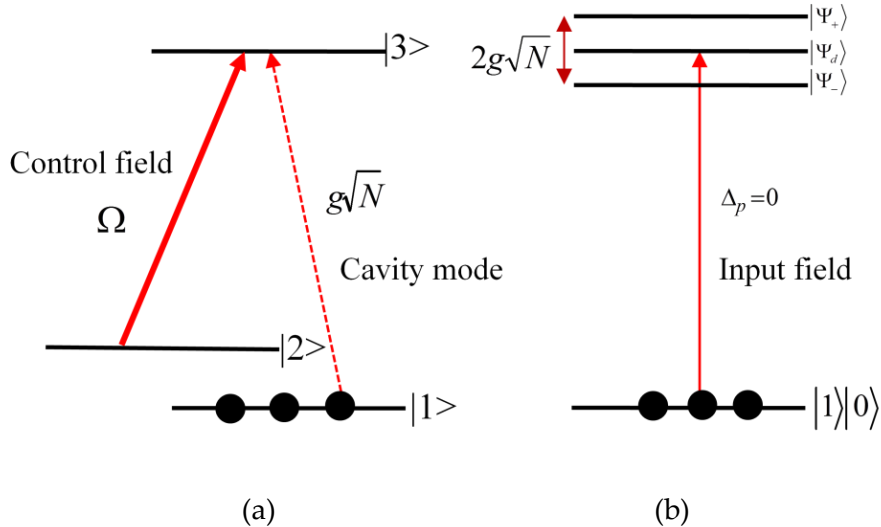

Fig. 3 Polariton picture of the coupled CQED system with $\Delta c=0$ and $\Delta=0$. (a) The three-level atoms are coupled by the cavity mode with the coupling coefficient $g\sqrt{N}$ and the free-space control laser with Rabi frequency $\Omega$. Three polariton states are created: two bright polariton states $|\Psi_+\rangle$ and $|\Psi_-\rangle$, and a dark polariton state $|\Psi_d\rangle$ (intra-cavity dark state) as shown in (b). (b) The input fields are coupled into the cavity. At $\Delta p=0$, it excites the intra-cavity dark state (cavity EIT) and leads to a near complete transmission of the input fields through the cavity.

III-2 Physical explanantion of cavity EIT switching

Fig. 3 provides a physical picture of the cavity EIT control of the perfect photon absorption in the CQED system. The composite CQED system and the control laser with Δc=0 and Δ=0 is depicted in Fig. 3(a) and has three first excited polariton states shown in Fig. 3(b). With $\Omega \ll g\sqrt{N}$, the two (bright) polariton states are $|\Psi_+\rangle = \frac{1}{\sqrt{2}}(|c\rangle|0_c\rangle + |a\rangle|1_c\rangle)$ and $|\Psi_-\rangle = \frac{1}{\sqrt{2}}(|c\rangle|0_c\rangle - |a\rangle|1_c\rangle)$ (($|1_c\rangle$ and $|0_c\rangle$ are one-photon and zero photon states of the cavity mode), and a dark polariton state (intra-cavity dark state) $|\Psi_d\rangle = \frac{1}{\sqrt{g^2 N + \Omega^2}}(g\sqrt{N}|1\rangle|1_c\rangle - \Omega|2\rangle|0\rangle) \approx |1\rangle|1_c\rangle$. When the input fields are coupled into the CQED system, it excites the two bright polariton states at $\Delta_p = \pm g\sqrt{N}$ and the dark polariton state at $\Delta_p$=0. The dark polariton state leads to a narrow transmission peak and is often referred to as cavity EIT [21-25].

Fig. 4 presents the calculated spectra for (a) the intra-cavity light intensity and (b) the output light intensity versus the frequency detuning of the input light field, Δp. It provides a spectral characterization of the cavity EIT control of the perfect photon absorption in the CQED system. Without the control laser (Fig. 4(a)), the input light fields are completely absorbed at Δp=0 and the photon energy is converted into the equal excitation

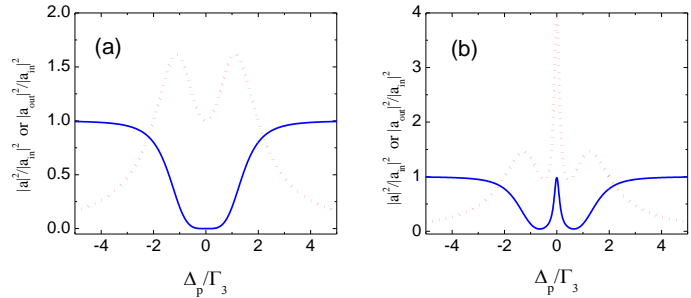

Fig. 4 With two input fields $a_{in}^l = a_{in}^r = a_{in}$, the output light intensity ($|a_{out}|^2 / |a_{in}|^2$) (blue solid curves) and the intra-cavity light intensity ($|a|^2 / |a_{in}|^2$) (red dotted curves) versus the frequency detuning of the input light $\Delta_p/\Gamma_3$. With $g\sqrt{N} = \kappa$, Δc=0, $\gamma_{12}$=0.001κ, and $\Gamma_3$=κ: (a) without the control laser (Ω=0), the perfect photon absorption occurs across $\Delta_p = 0$; (b) with the control laser present (Ω=0.2Γ and Δ=0), the perfect photon absorption disappears and the output field intensity equals to the input intensity $|a_{out}|^2 = |a_{in}|^2$.

of the two polariton states |Ψ+> and |Ψ-> (the resonance of the two polariton states are located at $\Delta_p = \pm g\sqrt{N} = \pm \kappa$). When the control laser is on (Fig. 4(b)), the cavity EIT is induced and a dark-polariton state (intra-cavity dark state) |Ψd> is created, which results in an even greater intra-cavity light field and a near perfect transmission of the input light field at Δp=0. We note that since the decoherence rate $\gamma_{12}$ is normally much smaller than $\Gamma_3$, the required control field intensity characterized by $\Omega^2 \gg 2\gamma_{12}\kappa$ for the high contrast control is in the weak field regime (below the saturation intensity). As an example, consider the D1 or D2 transitions of Rb atoms where $\Gamma_3$ ~ 6 MHz, $\gamma_{12}$ =0.0003$\Gamma_3$~0.002 MHz ($\gamma_{21}$≈10$^{-4}\Gamma_3$ has been observed in cold Rb atoms [27])), the required control Rabi frequency is $\Omega \gg \sqrt{2\gamma_{12}\kappa}$ ~ 0.2 MHz. Therefore, the high-contrast control of the perfect photon absorption with EIT can be done experimentally in the Rb atoms with a weak control

laser ($\Omega \ll \Gamma_3$) well below the saturation intensity, which is in the regime of nonlinear optics at low light intensities.

It is interesting to compare the cavity EIT control of the perfect photon absorption with the normal cavity EIT with identical system parameters ($\Delta c=0$, $\Delta=0$, $\Delta p=0$, $\kappa=\Gamma_3$, $g^2 N = \kappa\Gamma_3$, and $\Omega^2 \gg 2\gamma_{12}\kappa$) and observe the difference between the two cases. In the normal cavity EIT [21-25], there is only one input light field ($a_{in}^l \neq 0$ and $a_{in}^r = 0$, or vice versa): when the control field is not present ($\Omega=0$), the two output fields are $a^l = -\frac{a_{in}^l}{2}$ and $a^r = \frac{a_{in}^l}{2}$; when the control field is turned on ($\Omega \neq 0$), the two output fields becomes $a^l = 0$ and $a^r = a_{in}^l$ (the left output field is suppressed while the right output field is enhanced). The all-optical switching by the control field is inefficient and the switching contrast $|C| \approx 0.25$ for the left output field and $|C|=0.75$ for the right output field. Thus, for studies of all-optical switching, it is preferable to use the scheme of the cavity EIT control of the perfect photon absorption with two input fields.

## IV. SWITCHING OF CQED SYSTEM FROM PERFECT PHOTON ABSORBER TO NEAR PERFECT PHOTON REFLECTOR

In the strong collective-coupling regime, $g^2 N \gg \kappa\Gamma$, and under the condition of the perfect photon absorption, the input light photons are completely absorbed by the CQED system and the photon energy is converted into the excitation energy of one of the two polariton states with the input light frequency at $\Delta_p = g\sqrt{N}$ (or $\Delta_p = -g\sqrt{N}$) [28-29]. Thus there is no output light from the cavity and the CQED system acts as a perfect photon absorber. When the control laser is turned on and its frequency is tuned to the polariton resonance at $\Delta = g\sqrt{N}$ (or $\Delta = -g\sqrt{N}$), it creates the destructive quantum interference and suppresses the excitation of the polariton state, which eliminates the photon absorption by the CQED system and renders the CQED system as a perfect light reflector. With $\Delta = \pm g\sqrt{N}$, $\Delta_p = \pm g\sqrt{N}$, and $\Delta_c = 0$, and two identical input fields $a_{in}^r = a_{in}^l = a_{in}$ the intra-cavity light field is

$$a = \frac{2\sqrt{\kappa/\tau}(\Gamma_3 + \frac{\Omega^2}{\gamma_{12}} \mp ig\sqrt{N})a_{in}}{\kappa(\Gamma_3 + \frac{\Omega^2}{\gamma_{12}}) \mp ig\sqrt{N}(\Gamma_3 + \kappa + \frac{\Omega^2}{\gamma_{12}})}, \quad (15)$$

and the cavity output light field $a^l = a^r = a_{out}$ is

$$a_{out} = \frac{2\kappa(\Gamma_3 + \frac{\Omega^2}{\gamma_{12}} \mp ig\sqrt{N})a_{in}}{\kappa(\Gamma_3 + \frac{\Omega^2}{\gamma_{12}}) \mp ig\sqrt{N}(\Gamma_3 + \kappa + \frac{\Omega^2}{\gamma_{12}})} - a_{in} \quad (16)$$

In the strong collective-coupling regime, $g^2 N \gg \kappa\Gamma$, and under the condition of the perfect photon absorption ($\kappa=\Gamma_3$), when the control laser is off, the cavity output field is $a_{out}=0$ and the intra-cavity light field is $a = \frac{2a_{in}}{\sqrt{\kappa\tau}}$. That is, the two input light fields at $\Delta_p = \pm g\sqrt{N}$ are completely absorbed by the CQED system and the photon energies are converted into excited polariton state. The intra-cavity light field is at the peak value but there is no output light from the cavity. When the control laser with the frequency detuning $\Delta = \pm g\sqrt{N}$ is turned on, it creates the quantum interference that suppresses the polariton excitation [24] and turns off the perfect photon absorption. If the control laser Rabi frequency satisfies $\Omega^2 \gg \gamma_{21} g\sqrt{N}$, then the intra-cavity field is $a = 0$, and the cavity output field becomes $a_{out} \approx -a_{in}$. That is, by turning off or on of the control laser, the CQED system switches from a near perfect photon absorber to a near perfect photon reflector with a near perfect contrast $C \approx 1$.

IV-2 Physical explanation of the polariton interference switching

The physical picture for such interference control is depicted in Fig. 5. Under the conditions of $g^2 N \gg \kappa\Gamma$, $\Omega \ll g\sqrt{N}$, $\Delta c=0$, and $\Delta = g\sqrt{N}$, the control laser can be treated perturbatively. The cavity and atom coupling

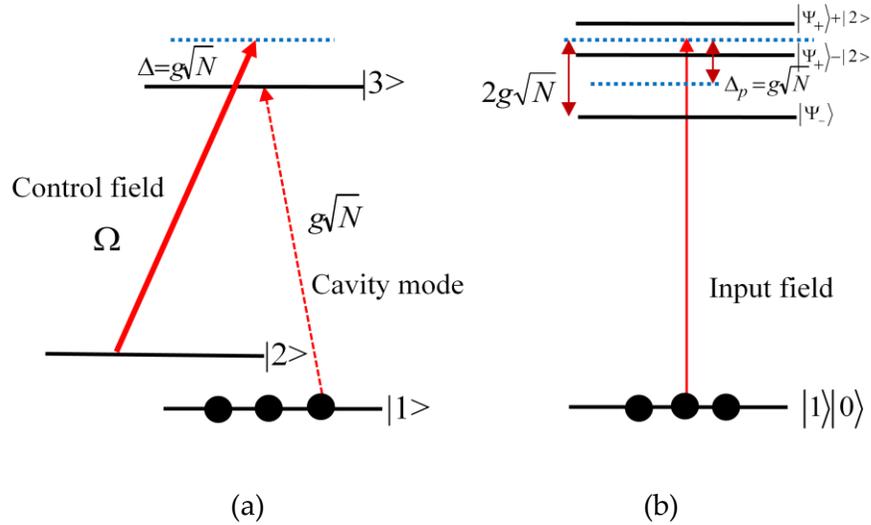

(a) (b)

Fig. 5 The polariton picture of the coupled CQED system. Without the control field, the input field is completely absorbed and excites the polariton state $|\Psi_+>$ at $\Delta_p = g\sqrt{N}$. When the control laser with $\Delta = g\sqrt{N}$ is turned on, a EIT-type interference is created and suppresses the polariton excitation, which results in the near perfect reflection of the input light field.

produces two first-excited polariton states $|\Psi_+> = \frac{1}{\sqrt{2}}(|c>|0_c> + |a>|1_c>)$ and $|\Psi_-> = \frac{1}{\sqrt{2}}(|c>|0_c> - |a>|1_c>)$, and then the control laser splits the polariton state $|\Psi_+>$ into two dressed polariton states $|\Phi_+> = \frac{1}{\sqrt{2}}(|\Psi_+> + |2>)$ and $|\Phi_-> = \frac{1}{\sqrt{2}}(|\Psi_+> - |2>)$, and creates two excitation paths for the input light fields as shown in Fig. 5(b). The EIT-type destructive interference between the two excitation paths suppresses the excitation of the polariton state at $\Delta_p = g\sqrt{N}$ and leads to the total reflection of the input light fields from the cavity. We note that the required conditions of $\Omega^2 \ll g^2 N$ and $\Omega^2 \gg \gamma_{21} g\sqrt{N}$ can be easily met in real atomic systems. For example, consider the D1 or

D2 transitions of Rb atoms where $\Gamma_3 \sim$ 6 MHz, $\gamma_{12} \sim 0.002$MHz, with $g\sqrt{N}=10\Gamma_3$ the required control Rabi frequency should satisfy $10\Gamma_3 \gg \Omega \gg 0.06\Gamma_3$, which can be easily met experimentally. Therefore, the high-contrast switching of the perfect photon absorption can be done experimentally in the Rb atoms with a weak control laser ($\Omega<\Gamma_3$) well below the saturation intensity.

Fig. 6(a) and 6(b) plot the normalized intensity of the output field $|a_{out}|^2/|a_{in}|^2$ and the normalized intensity of the intra-cavity field $|a|^2/|a_{in}|^2$ versus the input light frequency detuning. At the polariton resonance $\Delta_p = g\sqrt{N}$, the output fields $a^r = a^l = a_{out} = 0$, but the intra-cavity field $a$ is at the peak, indicating that the input fields are completely absorbed and the CQED system acts as a perfect photon absorber. Fig. 6(c) Fig. 6(d) plot the normalized intensity of the output field $|a_{out}|^2/|a_{in}|^2$ and the normalized intensity of the intra-cavity field $|a|^2/|a_{in}|^2$ versus the frequency detuning $\Delta_p/\Gamma_3$ when the control laser with $\Delta = g\sqrt{N}$ is on. It shows that at $\Delta_p = \Delta = g\sqrt{N}$, the output intensity $|a_{out}|^2/|a_{in}|^2 \approx 1$ and the intra-cavity field intensity $|a|^2/|a_{in}|^2 \approx 0$, indicating the no light can be coupled into the cavity and the input light fields are nearly completely reflected from the cavity mirrors.

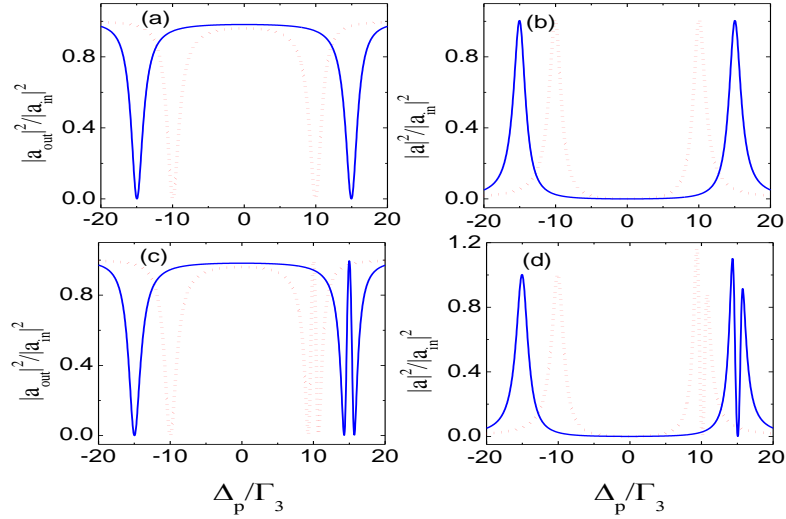

Fig. 6 With two identical input fields $a_{in}^l = a_{in}^r = a_{in}$ and $\Delta_c=0$, (a) the output field intensity $|a_{out}|^2/|a_{in}|^2$ and (b) the intra-cavity light intensity $|a|^2/|a_{in}|^2$ versus the input frequency detuning $\Delta_p/\Gamma_3$ when the control laser is off ($\Omega=0$). (c) The output field intensity $|a_{out}|^2/|a_{in}|^2$ and (d) the intra-cavity light intensity $|a|^2/|a_{in}|^2$ versus $\Delta_p/\Gamma$ when the control laser is present ($\Omega=0.8\Gamma_3$ and $\Delta = g\sqrt{N}$). $g\sqrt{N}=10\Gamma_3$ for red dotted curves and $g\sqrt{N}=10\Gamma_3$ for blue solid curves.

Next we compare the all-optical switching of the cavity output fields by the polariton interference control with the perfect photon absorption and without perfect photon absorption with identical system parameters

($\Delta c=0$, $\Delta_p = \Delta = g\sqrt{N}$, $\kappa=\Gamma_3$, $g^2N \gg \kappa\Gamma_3$, and $\Omega^2 \gg \gamma_{21}g\sqrt{N}$). In the all-optical switching by the polariton interference without the perfect photon absorption, there is only one input light field (say $a_{in}^l \neq 0$ and $a_{in}^r =0$) [30]: when the control field is not present ($\Omega=0$), the two output fields are $a^l = 0$ and $a^r = a_{in}^l$; when the control field is turned on ($\Omega\neq 0$), the two output fields becomes $a^l = -a_{in}^l$ and $a^r = 0$. That is, the two output fields are flip-flopped with each other by the control field and the switching is excecuted with a near perfect contrast C≈1. While for the CQED system with the perfect photon absorption, the two output fields are switched on or off simultaneously. Thus, the two schemes work with different outcomes and can be implemented for studies of all-optical switching requiring the all in-phase or opposite phase operations.

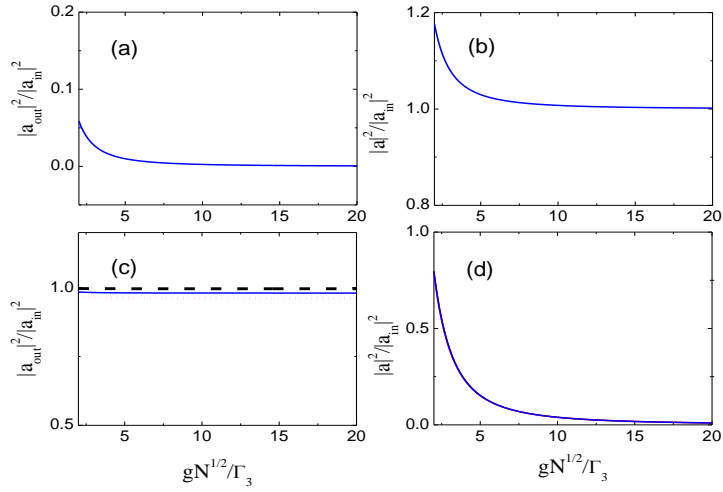

Fig. 7 (a) the output field intensity $|a_{out}|^2/|a_{in}^l|^2$ ($|a^r|^2 = |a^l|^2 = |a_{out}|^2$) and (b) the intra-cavity light intensity $|a|^2/|a_{in}^l|^2$ versus $g\sqrt{N}/\Gamma_3$ when the control laser is off ($\Omega=0$). (c) the output field intensity $|a_{out}|^2/|a_{in}^l|^2$ and (d) the intra-cavity light intensity $|a|^2/|a_{in}|^2$ versus $g\sqrt{N}/\Gamma_3$ when the control laser is present ($\Omega=0.5\Gamma_3$ and $\Delta = g\sqrt{N}$). In (c), $\gamma_{21}=0.001\Gamma_3$ for the black dashed line, $\gamma_{21}=0.005\Gamma_3$ for the blue solid line, and $\gamma_{21}=0.01\Gamma_3$ for the red dotted line. In (d), the black dashed line, the blue solid line, and the red dotted line with $\gamma_{21}$ values given above are nearly overlapped with each other. The other parameters are $\kappa=\Gamma_3$, $a_{in}^l = a_{in}^r$, $\Delta_p = \Delta = g\sqrt{N}$, and $\Delta_c=0$.

## V. NUMERICAL RESULTS FOR CONTROL FIELD WITH A POLARITON RESONANCE

To see how the performance of the CQED system varies versus the collective coupling coefficient $g\sqrt{N}$, we plot the output light intensity in Fig. 7(a) and the intra-cavity light intensity in Fig. 7(b) versus $g\sqrt{N}/\Gamma_3$ without the control laser ($\Omega=0$). It shows that when $g\sqrt{N} \gg \Gamma_3$, the CQED system acts as a perfect photon absorber: the input light is completely coupled into the cavity ($|a|^2 = |a_{in}^l|^2$), but there is no output light. When the control laser is present ($\Omega=0.5\Gamma_3$), and tuned to the polariton resonance ($\Delta=g\sqrt{N}$), the intra-cavity light field at $\Delta_p=g\sqrt{N}$ is near zero ($a=0$) and the input light is nearly completely reflected back from the cavity ($a_{out}\approx -a_{in}$). As the ground state decoherence rate $\gamma_{12}$ increases, the reflected light intensity decreases from unity, but the intra-cavity light intensity is maintained to be near zero, independent of $\gamma_{21}$ values.

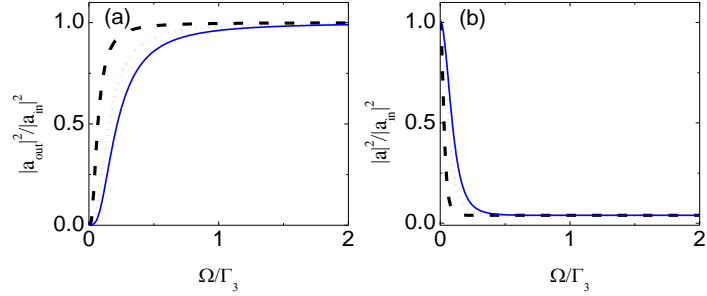

Fig. 8 With $\Delta_p=\Delta=g\sqrt{N}$ (both the control laser and the input fields are tuned to the polariton resonance), (a) the output field intensity $|a^r|^2=|a^l|^2=|a_{out}|^2$ and (b) the intra-cavity light intensity $|a|^2/|a^l_{in}|^2$ versus the control field Rabi frequency $\Omega/\Gamma_3$. $\gamma_{21}=0.001\Gamma_3$ for the black dashed line, $\gamma_{21}=0.005\Gamma_3$ for the red dotted line, and $\gamma_{21}=0.01\Gamma_3$ for the blue solid line. The other parameters are $g\sqrt{N}=10\Gamma_3$, $\kappa=\Gamma_3$, $a^l_{in}=a^r_{in}$, $\Delta_p=\Delta=g\sqrt{N}$, and $\Delta_c=0$.

Fig. 8(a) plots the output light intensity and Fig. 8(b) plots the intra-cavity light intensity versus the control Rabi frequency $\Omega/\Gamma_3$. It shows that with a control field near $\Omega\geq\Gamma_3$, the input light is essentially 100% reflected from the CQED system and no light can be coupled into the cavity. In particular, if the decoherence is small ($\gamma_{21}\leq 0.001\Gamma_3$), a weak control field $\Omega\ll\Gamma_3$ can be used to control the CQED system and convert the CQED system from a perfect absorber (no output light but with a large intra-cavity light field) into a near perfect reflector (near 100% of the reflection for the input light and no intra-cavity light) by turning off/on of the weak control field.

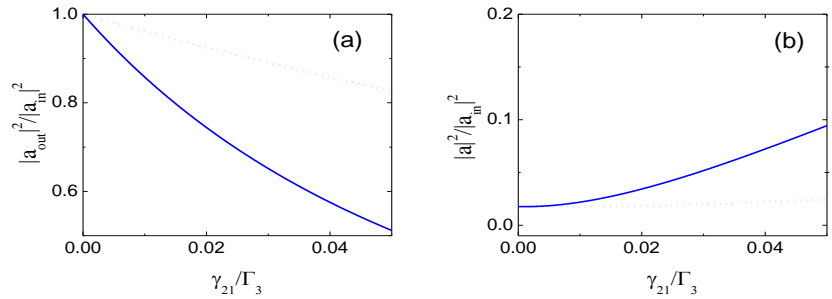

Fig. 9 (a) the output field intensity $|a_{out}|^2/|a_{in}|^2$ and (b) the intra-cavity light intensity $|a|^2/|a_{in}|^2$ versus the decoherence rate $\gamma_{21}/\Gamma_3$. The red dotted lines are for $\Omega=\Gamma_3$ and the blue solid lines are for $\Omega=0.5\Gamma_3$. The other parameters are $g\sqrt{N}=10\Gamma_3$, $\kappa=\Gamma_3$, $a^l_{in}=a^r_{in}$, $\Delta_p=\Delta=g\sqrt{N}$, and $\Delta_c=0$.

Fig. 9 plots (a) the output intensity and (b) the intra-cavity intensity versus the decoherence rate $\gamma_{21}$. It shows that manipulation of the CQED system with the control light is a coherent process: as $\gamma_{21}$ increases, the performance of the CQED system as a near perfect photon reflector degrades and the reflection coefficient $|a_{out}|^2/|a_{in}|^2$ decreases from unity (at the same time, more of the input light is coupled into the cavity). The calculations show that the performance of the CQED system as a reflector depends on the coupling Rabi frequency (a large $\Omega$ value results in a large reflectivity or better reflection performance), but is essentially independent of N, the number of atoms in the cavity when the CQED system is in the strong collective coupling regime ($g\sqrt{N} \gg \Gamma_3$).

## VI. CONCLUSION

In conclusion, we have shown that a CQED system with the perfect photon absorption can be understood in terms of an input light field resonantly coupled to an impedance matched, one-sided cavity, thus maximizing the light power transfer to the atomic excitation without any reflection loss. This is equivalent to an impedance matched AC circuit with the optimized load power. Thus it provides a new paradigm for the physics of coherent perfect photon absorption. We proposed and analyzed a scheme for the coherent control of the perfect photon absorption in a CQED system containing three-level atoms by the quantum interference induced by a control laser that couples the intra-cavity atoms from free space. The input light fields coupled into the cavity can be completely absorbed or transmitted by turning on or off of the control laser when the control laser creates the cavity EIT. By tuning the control laser to the polariton resonance and suppressing the polariton excitation, the CQED system can switch from being a perfect absorber to being a near perfect reflector. The CQED control scheme proposed and analyzed here can be applied to a variety of experimental CQED systems [19, 31-35] and may be useful for practical applications such as all-optical switching and optical multiplexing.

ACKNOWLEDGEMENT

Y. Zhu acknowledges support from the National Science Foundation under Grant No. 1205565.